\documentclass[10pt]{article}
\usepackage{euscript,amssymb,amsbsy}

\usepackage[dvips]{graphicx}
\catcode`\@=11
\renewcommand\footnoterule{%
  \kern-3\p@
  \hrule\@width.4\columnwidth
  \kern2.6\p@}
\renewcommand\@makefntext[1]{%
    \parindent 1em\noindent
    \hb@xt@1.8em{\hss$^{\@thefnmark}$)}\hspace{2pt}%
    \footnotesize\rmfamily#1} 
\def\@makefnmark{\hspace{.5pt}\hbox{$^{\@thefnmark}$%
\hspace{-1pt})}} 

\newcommand{\be}[1]{\begin{equation}\label{#1}}
\newcommand{\ee}{\end{equation}}
\newcommand{\ba}[1]{\begin{eqnarray}\label{#1}}
\newcommand{\ea}{\end{eqnarray}}
\newcommand{\rf}[1]{(\ref{#1})}
\newcommand{\nn}{\nonumber}
\newcommand{\tr}{\mbox{\rm tr}\,}

\def\RR{\mathbb{R}}

\def\CC{\mathbb{C}}

\def\PP{\mathbb{P}}

\def\ZZ{\mathbb{Z}}

\newcommand{\bsigma}{\boldsymbol{\sigma}}

\newcommand{\bvarphi}{\boldsymbol{\varphi}}

\newcommand{\bn}{\mathbf{n}}
\newcommand{\bp}{\mathbf{p}}

\newcommand{\cC}{\mathcal{C}}

\newcommand{\cF}{\mathcal{F}}
\newcommand{\cH}{\mathcal{H}}

\newcommand{\cK}{\mathcal{K}}
\newcommand{\cP}{\mathcal{P}}
\newcommand{\cT}{\mathcal{T}}

\newcommand{\fh}{\mathfrak{h}}

\newcommand{\fH}{\mathfrak{H}}

\newcommand{\tdE}{\tilde E}

\def\dg{\dagger}

\def\p{\partial}
\def\a{\alpha}
\def\b{\beta}
\def\d{\delta}

\def\sg{\sigma}
\def\e{\varepsilon}

\def\o{\omega}
\def\t{\theta}
\def\vfi{\varphi}

\def\ra{\rangle}
\def\la{\langle}

\renewcommand\Im{\mbox{Im}}

\newcommand{\dd}{\dagger}

\title{The $\cP\cT-$symmetric brachistochrone problem, Lorentz boosts  and non-unitary operator equivalence classes}

\author{Uwe G\"unther$^1$\footnote{e-mail:
u.guenther@fzd.de} and Boris F Samsonov$^{1,2}$\footnote{e-mail:
samsonov@phys.tsu.ru}\\[2ex]
$^1$ Research Center Dresden-Rossendorf,\\ POB 510119, D-01314
Dresden, Germany\\[1ex]
$^2$ Physics Department, Tomsk State University,\\ 36 Lenin Avenue,
634050 Tomsk, Russia\\[1ex]}
\date{20 July 2008}

\begin{document}

\maketitle
\begin{abstract} The $\cP\cT-$symmetric (PTS) quantum brachistochrone
problem is re-analyzed as quantum system consisting of a
non-Hermitian PTS component and a purely Hermitian component
simultaneously. Interpreting this specific setup as subsystem of a
larger Hermitian system, we find non-unitary operator equivalence
classes (conjugacy classes) as natural ingredient which contain at
least one Dirac-Hermitian representative. With the help of a
geometric analysis the compatibility of the vanishing passage time
solution of a PTS brachistochrone with the Anandan-Aharonov lower
bound for passage times of Hermitian brachistochrones is
demonstrated.

\end{abstract}

\section{Introduction\label{intro}}
Non-Hermitian $\cP\cT-$symmetric quantum mechanical (PTSQM) models
\cite{BB-PTSQM-1,cmb-rev} with exact $\cP\cT-$symmetry (PTS) and
diagonalizable spectral decomposition are known to be equivalent
to Hermitian quantum mechanical models \cite{ali-herm-1}. Under
the corresponding equivalence transformations non-Hermitian
Hamiltonians with differential expressions of local type are in
general mapped into strongly non-local Hermitian Hamiltonians
\cite{cmb-nonlocal,ali-nonlocal}, whereas non-Hermitian matrix
Hamiltonians are by conjugation mapped into Hermitian matrix
Hamiltonians. These equivalence relations led to the natural
conclusion \cite{ali-nonlocal} that PTSQM models with exact PTS
are a kind of economical writing of possibly complicated Hermitian
QM models --- and known properties of Hermitian QM will
straightforwardly extend to PTSQM in its exact symmetry sector.

In the recent consideration \cite{cmb-brach} a PTS quantum
brachistochrone model has been proposed which indicates on a
violation of the strict one-to-one equivalence PTSQM
$\Leftrightarrow$ standard QM. The model is of $2\times 2$ matrix
type, mathematically easily tractable and, therefore, it may serve
as toy model for physical concepts. Here we will show that the the
violation of the one-to-one equivalence follows from the fact that
the $\cP\cT-$symmetric brachistochrone model is built from
Hermitian operators and $\cP\cT-$symmetric (and therefore
non-Hermitian) operators simultaneously. The model is not
reducible to a setup with purely Hermitian operators. Rather the
Hermiticity of one component of the model will be connected with
the non-Hermiticity of another component, and vice versa. The
apparent physical inconsistency can be resolved by considering the
model as effective subsystem of a larger Hermitian system going in
this way beyond the one-to-one equivalence assumed, e.g., in
\cite{ali-brach-2007}.

Moreover we will find a hyperbolic structure underlying the
$\cP\cT-$symmetric model connected with the complex orthogonal
group $O(2,\CC)$ and indicating on certain structural analogies of
the $\cP\cT-$symmetric brachistochrone with Lorentz boosted spinor
systems. In this way it will appear natural to reconsider
$\cP\cT-$symmetric models connected by `boosts' as model families
and corresponding operators and observables as elements of
$O(2,\CC)$ conjugacy classes. In rough analogy to special
relativity we may introduce different reference frames. It turns
out that for $\cP\cT-$symmetric models of the type of
\cite{cmb-brach} the conjugacy classes contain at least one
Dirac-Hermitian operator. We find that the probabilistic contents
of models belonging to the same conjugacy class allows for a
natural interpretation as frame independence.

The basic subject of the present work is the PTSQM brachistochrone
problem of \cite{cmb-brach}. In section \ref{general}, we analyze
the equivalence relations between the representations of the
$\cP\cT-$symmetric system in terms of non-Hermitian and Hermitian
Hamiltonian. Re-parameterizing the mapping operator between
$\cP\cT-$symmetric and Hermitian Hamiltonian we show that it can
be re-interpreted as boost operator of a $2-$component spinor
setup. Using the close structural analogy to representations of
relativistic systems in different reference frames and the
representation invariance of the probabilistic content of the
model we introduce operator equivalence classes. In section
\ref{spin-flip} the passage time and probability content of the
brachistochrone are analyzed in detail. The underlying geometrical
structures of the brachistochrone problem are discussed in section
\ref{geometry} in terms of M\"obius transformations and
deformations of the Fubini-Study metric. They are visualized as
mapping between Bloch sphere setups and provide an explanation of
the vanishing passage time effect as geometric mapping artifact.
The results are summarized in the Conclusions (section
\ref{conclu}).

\section{\label{general} Operator equivalence classes}

\subsection{The $\cP\cT$-symmetric brachistochrone}

Let us briefly recall the quantum brachistochrone problem as
formulated in \cite{cmb-brach}. Given an initial state
$|\psi_i\ra$ and a final state $|\psi_f\ra$ of a quantum system
the problem consists in obtaining a PTS Hamiltonian $H$,
$[\cP\cT,H]=0$ which minimizes the time $t$ needed for the
evolution $U(t)=e^{-itH}:\ |\psi_i\ra\mapsto
|\psi_f\ra=U(t)|\psi_i\ra$. In \cite{cmb-brach} the Hamiltonian
$H$, the parity operator $\cP$ and the initial and final states
$|\psi_i\ra$ and $|\psi_f\ra$ were assumed as
\ba{0}
&&H=\left(%
\begin{array}{cc}
  re^{i\theta} & s \\
  s & re^{-i\theta} \\
\end{array}%
\right),\quad r,s,\theta\in\RR, \qquad \cP=\left(%
\begin{array}{cc}
  0 & 1 \\
  1 & 0 \\
\end{array}%
\right),\nn\\ &&|\psi_i\ra=(1,0)^T,\quad |\psi_f\ra=(0,1)^T.
\ea
The time inversion operator $\cT$ is antilinear and acts in the
present model as complex conjugation. The Hamiltonian $H$ has
eigenvalues $ E_\pm=r\cos(\theta)\pm\sqrt{s^2-r^2\sin^2(\theta)}$
so that exact $\cP\cT-$symmetry with $\Im E_\pm=0$ and
diagonalizability of $H$ hold for $s^2>r^2\sin^2(\theta)$.
Parameter configurations with $\t=0$ correspond to a purely
Hermitian (real symmetric) Hamiltonian, whereas configurations
with $s^2=r^2\sin^2(\theta)$ are related to the boundary between
exact and spontaneously broken PTS. These latter configurations
are characterized by coalescing eigenvalues
$E_+=E_-=E_0:=r\cos(\theta)$ and eigenvectors, lost
diagonalizability of $H\sim\left(%
\begin{array}{cc}
  E_0 & 1 \\
  0 & E_0 \\
\end{array}%
\right)$ and correspond to exceptional points \cite{GRS-EP-JPA}.
For fixed
\be{01}
\omega:=E_+-E_-
\ee
a Hamiltonian $H$ was found in
\cite{cmb-brach} which led to a vanishing evolution time $t=0$.

\subsection{Non-Hermitian Hamiltonian\label{non-Hermitian}}

As plausibly argued in \cite{ali-brach-2007}, a vanishing passage
time is impossible for a PTSQM model which by an equivalence
transformation can be one-to-one mapped into a purely Hermitian
$2\times 2$ matrix model. The apparent contradiction between the
results of \cite{cmb-brach} and \cite{ali-brach-2007} can be
resolved by noticing that the states $|\psi_i\ra$ and $|\psi_f\ra$
can be interpreted as eigenstates of a spin-$\frac12$ operator
$\sigma_z=\left(%
\begin{array}{cc}
  1 & 0 \\
  0 & -1 \\
\end{array}%
\right)=\sigma^\dd_z$ which is not $\cP\cT-$symmetric in the
representation \rf{0} for $\cP$. This means that the starting
assumptions of \cite{cmb-brach} (only $H$ is $\cP\cT-$symmetric)
and \cite{ali-brach-2007} (all the system is $\cP\cT-$symmetric)
are different and that therefore the conclusions are different.

Moreover, the approach of \cite{cmb-brach} implicitly indicates that
physical effects beyond the $2\times 2$ Hermitian matrix model can
be obtained from systems which comprise Hermitian and
$\cP\cT-$symmetric (non-Hermitian) subsystems simultaneously. For
this purpose it suffices to interpret the $\cP\cT-$symmetric
(non-Hermitian) components as dimensionally reduced (down-projected)
components of a larger Hermitian system.

Specifically, for the model \cite{cmb-brach} the non-Hermitian
$\cP\cT-$symmetric Hamiltonian $H$ in \rf{0} induces a non-unitary
evolution. This non-unitary evolution described by the Schr\"odinger
equation
\be{b0}
i\p_t \psi=H\psi
\ee
can be regarded as effective evolution in the dimensionally reduced
(down-projected) subsystem induced by the unitary evolution of the
larger closed system. Explicitly, the relation between the
down-projected and the closed system is easily demonstrated with the
help of a time independent Hermitian block matrix Hamiltonian $\hat
H=\hat H^\dd$ of the large system and its Schr\"odinger equation
\be{b1} i\p_t\hat \psi=\hat H\hat \psi
\ee
which takes the form
\be{b2}
i\p_t\left(%
\begin{array}{c}
   \psi \\
  \chi \\
\end{array}%
\right)=\left(%
\begin{array}{cc}
  A & B \\
  B^\dd  & D \\
\end{array}%
\right)\left(%
\begin{array}{c}
   \psi \\
  \chi \\
\end{array}%
\right).
\ee
Here $\chi$ denotes the wave function components in the subsystem
living in the Hilbert space components complementary to $\psi$. For
time-independent Hamiltonians $H$ and $\hat H$ the compatibility of
eqs \rf{b0} and \rf{b1} is ensured by a constraint on $H$ and the
matrix blocks $A=A^\dd$, $B$, $D=D^\dd$ in form of an algebraic
matrix Riccati equation
\be{b3}
H^2-(A+BDB^{-1})H-BB^\dd+BDB^{-1}A=0.
\ee
In general, this constraint is not invariant under Hermitian
conjugation so that accordingly $H$ is, in general, non-Hermitian.
The effect of the non-unitary evolution of the down-projection is
easily understood by noticing that vectors $\hat \psi$, $\hat\phi$
orthogonal in a large (Hilbert) space $\hat \cH=\cH_1\oplus
\cH_2\ni \hat\psi,\hat\phi$ remain orthogonal under unitary
evolution in this space. Their down-projected components are, in
general, non-orthogonal  in the lower-dimensional subspaces
\ba{b4}
\la
\hat\psi|\hat\phi\ra_{\hat\cH}=(\psi_1,\phi_1)_{\cH_1}+(\psi_2,\phi_2)_{\cH_2}&=&0\nn\\
(\psi_{1},\phi_{1})_{\cH_{1}}=-(\psi_{2},\phi_{2})_{\cH_{2}}&\neq
&0
\ea
and evolve in these subspaces non-unitarily.

Further on, we restrict our attention to the effective
down-projected system with non-Hermitian $\cP\cT-$sym\-me\-tric
Hamiltonian $H$ whose eigenvalues are purely real (sector of exact
$\cP\cT-$symmetry). In contrast to a Hermitian Hamiltonian the
eigenvectors of $H$ are in general non-orthogonal in Hilbert space
and, therefore, $H$ is not a von-Neumann observable. In this
regard it should be noted that in modern quantum theory the
concepts of ``observable'' and ``measurement'' are understood in a
wider sense than in the early times of Bohr, von Neumann, Dirac et
al. In particular, one does not associate Hermiticity with a
necessary attribute of an observable \cite{Holevo} anymore.
Non-orthogonal vector sets  appear naturally after measurements of
observables. They are used in constructing non-orthogonal
decompositions of the identity operator, so called positive
operator valued measures (POVMs), and provide a consistent
probabilistic interpretation of the measurement process
\cite{Holevo,Peres,Nielsen,QI,diosi}. The corresponding approach
is one of the cornerstones of quantum information and computation
theory \cite{Nielsen,QI,diosi}. Von Neumann observables and their
orthogonal projector decompositions of the identity operator are
connected with repeatable measurements and provide sharp
observables \cite{diosi}, whereas ``generalized observables'' with
non-orthogonal identity decompositions are unsharp (smeared)
observables and correspond to nonrepeatable, purely probabilistic
measurements \cite{diosi}.

\subsection{Lorentz boost analogy\label{boost}}

Let us recall a few basic facts on the $\cP\cT-$symmetric
Hamiltonian $H$ in \rf{0}. This Hamiltonian is selfadjoint with
regard to the indefinite $\cP\cT$ inner product
$\cP\cT|E_k\ra\cdot|E_l\ra$ (see \cite{krein-pts}) and it is
therefore selfadjoint in the Krein space\footnote{A Krein space is
a Hilbert space endowed with an additional indefinite inner
product structure \cite{azizov,L2}.} $(\cK_\cP,[.,.]_\cP)$,
$\cK_\cP\cong \CC^2$ with the indefinite metric defined by  the
parity inversion $\cP$ as $[.,.]_\cP:=\la .|\cP|.\ra$. Moreover,
there exists a Hermitian operator $\eta=\eta^\dd >0$ so that
\be{b5}
\eta H=H^\dd\eta
\ee
and, hence, that $H$ is self-adjoint (quasi-Hermitian in the sense
of \cite{hahne}) in the Hilbert space $(\cH_\eta,
\la.,.\ra_\eta)$, $\cH_\eta\cong \CC^2$ endowed with $\eta $ as
positive definite metric $\la.,.\ra_\eta:=\la .|\eta|.\ra$
\cite{ali-pseudo-herm}. Identifying the $\cC\cP\cT$ inner product
(see, e.g. \cite{cmb-rev} ) with this $\eta-$defined inner product
one finds $\cC\cP\cT|E_k\ra\cdot|E_l\ra=\la
E_k|(\cC\cP)^T|E_l\ra=\la E_k|\eta|E_l\ra$ and, hence,
$\eta^T=\cC\cP$. Together with the relation $\eta^{-1}=\cC\cP$
obtained in \cite{ali-eta-JMP} this implies $\eta^T=\eta^{-1}$.
For general $N\times N-$matrix models this means that $\eta$ is an
element of the complex orthogonal group $O(N,\CC)\ni \eta$ [and
$\eta\in SO(N,\CC)$ in case of $\det(\eta)=1$] additionally to the
Hermiticity $\eta=\eta^\dd$.

In contrast to Hermitian Hamiltonians, the spectrum of
$\cP\cT-$symmetric Hamiltonians may consist of real eigenvalues as
well as of complex conjugate eigenvalue pairs. Concerning the
brachistochrone problem  we restrict our attention to
$\cP\cT-$symmetric Hamiltonians with purely real spectrum. Such
Hamiltonians $H$ are known to be in a one-to-one equivalence
relation to Hermitian Hamiltonians $h$ \cite{ali-herm-1}
\be{b5a}
\rho:\ H\mapsto h=\rho H \rho^{-1},\quad h=h^\dd,
\ee
where due to \rf{b5} it holds $\eta=\rho^\dd\rho$. Obviously, up to
a unitary transformation of $h$ one may set $\rho=\rho^\dd$ so that
\be{b6}
\eta=\rho^2
\ee
and $\rho$ itself is a complex orthogonal rotation as well: $\rho\in
O(N,\CC)$.

Let us now explicitly apply these considerations to the
brachistochrone system of \cite{cmb-brach}. For this purpose we
represent the Hamiltonian \rf{0} as
\be{h1}
H=a_0I_2+s\left(%
\begin{array}{cc}
  i\sin(\a) & 1 \\
  1 & -i\sin(\a) \\
\end{array}%
\right),\qquad \sin(\a):=\frac rs \sin(\t),\quad a_0:=r\cos(\t).
\ee
Its bi-orthogonal non-normalized eigenvectors have the form
\cite{GRS-EP-JPA}
\ba{h2}
&&|E_\pm\ra=c_\pm \chi_\pm\,,\qquad |\tilde
E_\pm\ra=d_\pm^*\chi_\pm^*\,,\quad c_\pm,d_\pm\in\CC^*,\nn\\
&&\chi_\pm:=\left(1,
  -i\sin(\a)\pm\sqrt{1-\sin^2(\a)}
\right)^T
\ea
and it holds $\la\tilde E_\mp|E_\pm\ra=0$ $\forall \a$, and
$\la\tilde E_\pm|E_\pm\ra=c_\pm d_\pm \chi_\pm^T\chi_\pm\neq 0$
$\forall \a\neq (N+1/2)\pi$, $N\in\ZZ$. The values $\a= (N+1/2)\pi$
correspond to exceptional points (EPs) of the spectrum
\cite{GRS-EP-JPA} and the eigenvectors due to
$\chi(\a=\pm\pi/2)=(1,\mp i)^T$ become isotropic (self-orthogonal)
$\chi_\pm^T\chi_\pm=0$ at these points. In \cite{GRS-EP-JPA} several
arguments have been listed which indicate on a strong similarity of
these isotropic eigenvectors and the isotropic light-like vectors
well known from special relativity. Here, we take this analogy
literally and conjecture the ansatz $\sin(\a)=v/c$ so that
$\chi_\pm$ contains terms which disappear in the light-cone limit
$|v|\to c$ in the typical relativistic way
\be{h3}
\chi_\pm=\left( 1,
  -i\frac vc\pm\sqrt{1-\frac{v^2}{c^2}}
\right)^T\,.
\ee
On its turn this suggests the usual reparametrization
\be{h3a}
\sin(\a)=v/c=:\tanh(\b)
\ee
with (see \rf{01} and \rf{h1})
\be{om}
\cosh(\b)=\frac{2s}{\omega}\,.
\ee

From $\cP$ in \rf{0} and the operator $\cC$ (see, e.g.
\cite{cmb-rev}) which encodes the dynamical mapping between the
Krein-space $\cP\cT$ inner product and the Hilbert space $\cC\cP\cT$
inner product we find the explicit representation of the metric
\ba{q1}
\eta=\cP\cC&=&\frac1{\cos(\a)}\left(%
\begin{array}{cc}
  1 & -i\sin(\a) \\
  i\sin(\a) & 1 \\
\end{array}%
\right),\qquad \det(\eta)=1,\qquad \eta\in SO(2,\CC)\nn\\
&=&\left(%
\begin{array}{cc}
  \cosh(\b) & -i\sinh(\b) \\
  i\sinh(\b) & \cosh(\b) \\
\end{array}%
\right)=e^{\b\sigma_y},\qquad \sigma_y=\left(%
\begin{array}{cc}
  0 & -i \\
  i & 0 \\
\end{array}%
\right)
\ea
and with \rf{b6} the transformation $\rho:\ H\mapsto h$
\be{q2}
\rho=e^{\b\sigma_y/2}=\left(%
\begin{array}{cc}
  \cosh(\b/2) & -i\sinh(\b/2) \\
  i\sinh(\b/2) & \cosh(\b/2) \\
\end{array}%
\right)\in SO(2,\CC)\,.
\ee
Due to $\rho^\dd \sg_z\rho=\sg_z$ the transformation is
pseudo-unitary $\rho\in SU(1,1)$ as well. In terms of the
$\b-$parametrization the Hamiltonian $H$ can be represented via
\rf{h1}, \rf{h3a} and \rf{om}, i.e. via $(r,s,\t)\mapsto
(r,\o,\b)$, as
\be{q10}
H(\b)= a_0I_2+\frac\o2\left(%
\begin{array}{cc}
  i\sinh(\b) & \cosh(\b) \\
  \cosh(\b) & -i\sinh(\b) \\
\end{array}%
\right),\qquad \qquad a_0=r\cos(\t)=\sqrt{r^2-\frac{\o^2}
4\sinh^2(\b)}.
\ee
Its Hermitian equivalent takes the form
\be{q11}
h=\rho H\rho^{-1}=a_0 I_2+\frac\o2\sigma_x\,.
\ee
The transformation invariant energy offset $a_0 I_2$ produces a
general phase factor of the wave function and, for tuned
$r=\frac\o2 \cosh(\b)$, takes the $\b-$independent value
$a_0=\o/2$. The dynamically relevant non-trivial matrix terms of
the Hamitlonians $H$ and $h$ show a certain structural similarity
with the chiral components of the Dirac equation in its Weyl
representation \cite{ryder}
\be{q12} D_W\Psi\equiv\left(
  \begin{array}{cc}
    -m & p_0+\bsigma\cdot\bp \\
    p_0-\bsigma\cdot\bp & -m \\
  \end{array}
\right)\left(
         \begin{array}{c}
           \phi_R(\bp) \\
           \phi_L(\bp) \\
         \end{array}
       \right)=0
\ee
\be{q13}
\phi_{R,L}(\bp):=e^{\pm \bvarphi\cdot\bsigma/2}\phi_{R,L}(0)\,.
\ee
Here $\phi_{R,L}(\bp)$ denote the chiral right and left 2-component
spinors of a spin-$\frac12$ particle with energy $p_0$, rest mass
$m$ and momentum $\bp$ directed along the unit vector $\bn$,
\be{q13a}
p_0=m\,\cosh(\varphi),\qquad \bp=\bn m\,\sinh(\varphi),
\ee
$\phi_R(0)=\phi_L(0)$ are the corresponding rest frame chiral
spinors and $e^{\pm \bvarphi\cdot\bsigma/2}=
\cosh(\varphi/2)I_2\pm \bsigma\cdot\bn\sinh(\varphi/2)$ are the
pure boosts relating the spinors in the two frames.

With the help of the rotation
\be{q14a}
V=\frac1{\sqrt2}(I_2-i\sg_y)\in SU(2)
\ee
we find from \rf{q11}
\be{q14}
\fh:= V^{-1}[h-a_0I_2]V=m \sg_z,\quad m:=\frac{\omega}{2},\quad
\ee
and from \rf{q10}
\be{q15}
\fH(\b):=V^{-1}[H(\b)-a_0I_2]V =\rho^{-1}\hat h\rho =
\sg_z(p_0+\sg_y p_y)=\sg_z(p_0+\bsigma\cdot\bp)
\ee
with
\be{q14a2}
p_0:=m\,\cosh(\b)\,,\quad p_y:=m\,\sinh(\b)\,,
\ee
 $p_x=p_z=0$
and $\fH(-\b)=\sg_z(p_0-\bsigma\cdot\bp)$ so that
\be{q16}
\Sigma_z D_W\psi=\left(
                 \begin{array}{cc}
                   -\sg_z m & \fH(\b) \\
                   \fH(-\b) & -\sg_z m \\
                 \end{array}
               \right)\left(
                        \begin{array}{c}
                          \phi_R(p_y) \\
                          \phi_L(p_y) \\
                        \end{array}
                      \right)=0,\qquad \Sigma_z:=\left(
                                                 \begin{array}{cc}
                                                   \sg_z & 0 \\
                                                   0 & \sg_z \\
                                                 \end{array}
                                               \right)=I_2\otimes
                                               \sg_z
\ee
with $\phi_{R,L}(p_y)=e^{\pm \b\sg_y/2}\phi_{R,L}(0)$. This means
that $h$ and $\phi$ can be related via the chiral components
$\phi_R(0)=\phi_L(0)=V^{-1}\phi$ to a massive spin-$\frac12$
particle (with rest mass $m=\omega/2$) in its rest frame
(co-moving frame). In contrast,  $H(\b)$ and $\psi$ can be
associated to the same particle observed from a Lorentz boosted
frame (laboratory/observer frame) \cite{ryder}. Energy and
momentum are, as usual, related by the mass shell condition
$p_0^2-p_y^2=m^2$, which guaranties the compatibility of the
system \rf{q16}. The transformation $\rho(\b)=e^{\b\sg_y/2}$ is
then the usual Lorentz boost acting in the 2-component spinor
representation\footnote{We note that $\sg_y\in so(2,\CC)$ (with
$\sg_y=-\sg_y^T$) belongs also to $su(1,1)$ (due to $\sg_y=-\mu
\sg_y^\dd\mu$ with indefinite $\mu=\cP=\sg_x$ from \rf{0} as well
as with diagonal $\mu=\sg_z$).}.

\subsection{Operator equivalence classes and their statistical content\label{op-equiv}}

The structural analogy of the $\cP\cT-$symmetric matrix system and
the chiral components of relativistic particle systems leads to the
following natural assumption. Similar as physical observables in
relativistic systems can be measured in different reference frames,
we can associate PTSQM systems represented in terms of Hamiltonians
$H$ and $h$ with different physical reference frames. In analogy
with relativistic systems where the observables described in
different reference frames are related by Lorentz transformations
and can be associated to orbit sections of unitary representations
of the Lorentz group one can assume for PTSQM systems that the
corresponding observables in different reference frames are related
by equivalence transformations of the complex orthogonal group
$SO(N,\CC)$. In accordance with subsection \ref{non-Hermitian}, the
fact that for the PTSQM system the corresponding transformations are
not unitary but of $SO(N,\CC)$ type can be attributed to the
projection of the larger unitary system to its lower dimensional
subsystem\footnote{Mathematically, this follows trivially from the
fact that a similarity transformation between two complex symmetric
matrices is necessarily a complex orthogonal rotation.}.

For the brachistochrone model \cite{cmb-brach} we have one frame
$\cF_1$ associated with the non-Hermitian Hamiltonian $H$ and the
Hermitian spin operator $S_z$. The equivalence transformation
$\rho$ maps $H$ into the Hermitian Hamiltonian $h=\rho H\rho^{-1}$
which can be associated with a second reference frame $\cF_2$.
Simultaneously with $H$ the spin operator $S_z$ maps into
\ba{q17}
S_z\mapsto s_z=\rho S_z\rho^{-1}
\ea
which due to the non-unitarity of $\rho(\b\neq 0)$, i.e.
$\rho^\dd=\rho\neq \rho^{-1}$, is non-Hermitian $s_z\neq
s_z^\dd=\rho^{-1} S_z\rho$. Hence, in both frames $\cF_1$ and
$\cF_2$ the system is described in terms of Dirac-Hermitian as
well as non-Dirac-Hermitian operators. Therefore it cannot be
regarded as fundamental in the sense that there exists a frame
where all operators are Dirac-Hermitian simultaneously and where
they can be treated as von-Neumann observables.

In usual quantum mechanics, the expectation values of the physical
observables as well as the probabilities to observe them are
invariant under unitary transformations of the operators and state
vectors so that the physics is not affected by the concrete
representation. Below we show that the probabilities and
expectation values of an observable are independent of the
representation used to calculate them. In particular, this means
that the properties of a given observable can be calculated in a
representation (or ``reference frame'') where the observable is
associated with a Hermitian operator. Therefore the usual quantum
mechanical orthogonal measurements may be applied to this
observable in the specific frame. Using this property we show that
both probabilities and average values can also be calculated in a
representation where the observable is described by a
non-Hermitian operator (``non-Hermitian'' or ``observer reference
frame''). This leads to a generalization of the notions of both
the ``statistical operator'' describing the state of a quantum
object and the projection operators on eigenstates of the
observable.

Let a quantum system be in  pure state $|\vfi\ra$,
$\la\vfi|\vfi\ra=1$, and we wish to measure an observable (energy)
described by a Dirac-Hermitian operator $h=h^\dg$. Then according
to the axioms of  standard (von Neumann) quantum mechanics we can
detect only eigenvalues $E_i$ of $h$,
\be{m1}
h|e_i\ra=E_i|e_i\ra\,,\quad \la e_i|e_j\ra=\d_{ij}
\ee
(for simplicity we assume $h$ acting in a
finite-dimensional Hilbert space) with the probabilities
\be{m2}
p_i=|\la e_i|\vfi\ra|^2=\mbox{Tr}(P_i\varrho)\,,\quad
P_i=|e_i\ra\la e_i|\,,\quad
\varrho=|\vfi\ra\la\vfi|\,.
\ee
If the value $E_k$ appeared as a measurement result then after the
measurement the state of the system is described by the state
vector (up to an unessential normalization factor) $|e_k\ra$. If
now we change the representation (``reference frame'') using a
non-unitary nonsingular similarity transformation (see \rf{q17})
\be{m3}
|\vfi\ra=\rho|\psi\ra=\rho^{-1}|\tilde\psi\ra\,,\quad
|e_i\ra=\rho|E_i\ra=\rho^{-1}|\tdE_i\ra\,,\quad
h=\rho H\rho^{-1}=\rho^{-1}H^\dg\rho
\ee
with $|E_i\ra$ and $|\tdE_i\ra=\rho^2|E_i\ra$ being eigenvectors of $H$
and $H^\dg$ respectively,
\be{m4}
H|E_i\ra=E_i|E_i\ra\,,\quad H|\tdE_i\ra=E_i|\tdE_i\ra
\ee
then
\be{pi}
p_i=
\la\tilde\psi|E_i\ra\la \tdE_i|\psi\ra=
\mbox{Tr}(\Pi_i\Upsilon)=
\la\psi|\tdE_i\ra\la E_i|\tilde\psi\ra=
\mbox{tr}(\Upsilon^\dg\Pi_i^\dg)
\ee
where
\be{so}
\Pi_i=|E_i\ra\la\tdE_i|\,,\quad
\Upsilon=|\psi\ra\la\tilde\psi|\,.
\ee
The expectation value of the energy can be expressed in terms of
$|E_i\ra$, $\tdE_i\ra$ and $|\psi\ra$, $|\tilde\psi\ra$ as well
\be{m5}
\la E\ra=\la\vfi|h|\vfi\ra=\mbox{Tr}(h\varrho)=
\sum_{i=1}^NE_i\la\tilde\psi|E_i\ra\la\tdE_i|\psi\ra=
\la\tilde\psi|H\psi\ra=\mbox{Tr}(H\Upsilon) =
\la\psi|H^\dg\tilde\psi\ra=\mbox{Tr}(\Upsilon^\dg H^\dg) \,.
\ee
Thus, the operators $\Upsilon$, $\Upsilon^\dg$ play the role of
statistical operators for a pure state associated with the vector
$|\psi\ra$, whereas $\Pi_k$, $\Pi_k^\dg$ describe the observable
corresponding to the Hamiltonian (energy) of the system in these
``non-Hermitian frames''. The state with definite value of the
energy is described either by the quasi-projectors
$\Pi_i=|E_i\ra\la\tdE_i|$ or by $\Pi_i^\dg$.

If the outcome $E_k$ is detected as a measurement result then the
state of the system after the measurement (up to a normalization
factor) in the Hermitian frame  is described by the vector
$P_k|\psi\ra\sim |e_k\ra$ and in non-Hermitian frames by either
$|E_k\ra=\rho^{-1}|e_k\ra$ or $|\tilde E_k\ra=\rho|e_k\ra$ with
statistical operators $\Upsilon_k=|E_k\ra\la\tilde E_k|$ and
$\Upsilon_k^\dg$.

Unitary equivalent classes,  where the same physical observables
are represented by different operators related with unitary
transformations, are, evidently, subclasses of these more general
equivalence transformations.

Our final comment here is that the probabilities \rf{pi} are
intimately related with experiments on unambiguous state
discrimination which, in turn, are based on generalized
observables, POVM and Naimark's dilation (extension) theorem.
Moreover, quasi-projectors \rf{so} appear in a natural way when an
observable related with a specific symmetry operator in an
extended space is measured \cite{Our2}.

\section{Spin-flips under a non-Hermitian evolution\label{spin-flip}}

Let us illustrate this scheme by re-analyzing the
$\cP\cT-$symmetric quantum brachistochrone problem of
\cite{cmb-brach} as pseudo-unitary evolution (spin-flip) problem
of a Hermitian spin-$\frac12$ observable. According to the
equivalence relations found in section \ref{op-equiv} there are
two equivalent ways to calculate the spin-flip probabilities. One
may either consider the pair $S_z,H$ with $S_z:=\sg_z=S_z^\dd$,
$H\neq H^\dd$ and find the pseudo-unitary evolution
operator%
\footnote{From a time independent diagonalizable $H$ with $\cP H=
H^\dd \cP$, $\eta H= H^\dd\eta$ and purely real spectrum an
evolution operator $U(t)=e^{-itH}$ can be constructed which
fulfills $\la U(t)\psi|\cP|U(t)\chi\ra=\la \psi|\cP|\chi\ra$ as
well as $\la U(t)\psi|\eta|U(t)\chi\ra=\la \psi|\eta|\chi\ra$ so
that $U(t)$ is $\cP-$pseudo-unitary in the Krein space
$(\cK_\cP,\la.|\cP|.\ra)$ and $\eta-$pseudo-unitary in the Hilbert
space $(\cH_\eta,\la.|\eta|.\ra)$.} $U(t)=e^{-itH}$ acting on the
spin eigenstates $|\uparrow\ra$, $|\downarrow\ra$ of $S_z$. Or,
alternatively, one may consider the equivalent pair $s_z,h$
consisting of a non-Hermitian operator $s_z\neq s_z^\dd$ whose
eigenstates undergo a unitary evolution $u(t)=e^{-ith}$ governed
by $h=h^\dd$. We choose the first way of calculation (following
\cite{cmb-brach}) and obtain $U(t)$ via exponentiation of the
$\cP\cT-$symmetric $2\times 2-$matrix Hamiltonian \rf{0} as
\ba{UB}
U(t)&=&e^{-i Ht}=\sum_ke^{-i
E_kt}|\psi_k(0)\ra\la\psi_k(0)|\nn\\
&=&\frac{e^{-irt\cos\t}}{\cos \a} \left(
\begin{array}{cc}
\cos(\frac{\o t}{2}-\a) & -i\sin(\frac{\o t}{2})\\
-i\sin(\frac{\o t}{2}) & \cos(\frac{\o t}{2}+\a)
\end{array}
\right) \ne U^\dd(t)
\ea
where $\sin\a:=\frac rs\sin\theta$ and $\o:=2s|\cos\a|=\Delta E$ is
the difference of eigenvalues of $H$. Applying $U(t)$ to the initial
spin-up state, $|\uparrow\ra$, we reproduce the previously reported
result of \cite{cmb-brach}
 \be{psit}
|\psi(t)\rangle=\frac{e^{-irt\cos\t}}{\cos\a} \left(
\begin{array}{c}
\cos(\frac{\o t}{2}-\a) \\ -i\sin\frac{\o t}{2}
\end{array}
\right).
 \ee
The probabilities to find the spin either up or down at any time
moment $t>0$ for a system being in the state \rf{psit} are
calculated using the usual quantum mechanical prescriptions in the
Hilbert space $(\cH,\la.|.\ra)$ (measurement of $S=\sg_z$)
\be{pud}
p_{\uparrow}(t)=
\frac{\la\psi(t)|\uparrow\ra\la\uparrow|\psi(t)\ra}%
{\la\psi(t)|\psi(t)\ra}\,,\qquad p_{\downarrow}(t)=
\frac{\la\psi(t)|\downarrow\ra\la\downarrow|\psi(t)\ra}%
{\la\psi(t)|\psi(t)\ra}
\ee
and give in the present case
\be{p}
p_{\uparrow}=\frac{\cos^2\left(\frac{\omega
t}2-\a\right)}{\cos^2\left(\frac{\omega
t}2-\a\right)+\sin^2\left(\frac{\omega t}2\right)}\,,\qquad
p_{\downarrow}=\frac{\sin^2\left(\frac{\omega
t}2\right)}{\cos^2\left(\frac{\omega
t}2-\a\right)+\sin^2\left(\frac{\omega t}2\right)}\,.
\ee
From here we find the time intervals
\be{m7}
\Delta t_{\uparrow\to\downarrow}=\frac{\pi+2 \a}{\Delta
E}\,,\qquad \Delta t_{\downarrow\to\uparrow}=\frac{\pi-2
\a}{\Delta E}
\ee
necessary for the first spin flips from up to down and back
respectively. For all values $\a\in[-\pi/2,0)$ the evolution time
lies below the Anandan-Aharonov lower bound $\Delta
t_{\uparrow\to\downarrow}\ge \frac{\pi}{\Delta E}=:\Delta_{AA}$
for a spin-flip evolution in a Hermitian system
\cite{anandan-aharonov-prl-1990}. In the special case
$\a\to-\pi/2$ with $\Delta E$ fixed the zero-passage time result
$\Delta t_{\uparrow\to\downarrow}(\a\to-\pi/2)\to 0$ from
\cite{cmb-brach} is reproduced. In \cite{GRS-EP-JPA} this regime
has been related to an exceptional point of the spectrum of $H$
where its two eigenvectors coalesce so that the Hilbert space
distance between them vanishes. Subsequently we show that in the
equivalent system with Hermitian $h=h^\dd$ the originally
orthogonal $|\psi_i\ra$, $|\psi_f\ra\in \cH$ in the Hilbert space
$\cH$ are mapped into nearly coalescing $|\phi_i\ra,|\phi_f\ra\in
\tilde \cH$ so that $\Delta
t_{\uparrow\to\downarrow}(\a\to-\pi/2)\to 0$ is not connected with
a violation of the Anandan-Aharonov lower bound $\Delta
t_{\uparrow\to\downarrow}\ge \frac{\pi}{\Delta E}$. Rather it can
be attributed to changes in the Hilbert space metric induced by
the mapping $\rho: \cH\to \tilde \cH$.

Before we turn to the corresponding geometrical considerations two
comments are in order:\\
{\bf 1.} The total time for a spin-flip followed by a flip back
$\Delta t_{\uparrow\to\downarrow\to\uparrow}$ remains invariant
$\Delta t_{\uparrow\to\downarrow\to\uparrow}=\frac{2\pi}{\Delta E}$
independently of the non-Hermiticity parameter $\a$  --- a result
obtained recently also in \cite{giri-brach}.\\
{\bf 2.} With regard to the concept of an observable as conjugacy
orbit the general solution technique for a PTS brachistochrone
problem will comprise the following steps. (i) Given a family of
diagonalizable PTS Hamiltonians $H$ one finds the transformations
$\rho$ which render them Hermitian in the Hilbert space
$\tilde\cH$. (ii) Initial and final state $|\psi_{i,f}\ra\in\cH$
are to be mapped into $|\phi_{i,f}\ra\in\tilde\cH$. (iii) With
these vectors as initial and final states the brachistochrone
problem is solved along standard techniques for Hermitian
Hamiltonians \cite{brody-1} singling out a specific Hermitian
Hamiltonian  $h_b$. (iv) Its non-Hermitian representative $H_b$
from the same conjugacy orbit is the solution of the PTS
brachistochrone problem.

\section{Geometry of the $\cP\cT-$symmetric brachistochrone\label{geometry}}

The origin of the zero-passage time solution of the
$\cP\cT-$symmetric brachistochrone problem is easily understood by
studying the geometric properties of the $\eta-$related mapping
$\rho$ and its action on the projective Hilbert (state) space of
the model $\CC\PP^1\cong (\CC^2-\{0\})/\CC^*\cong \hat\CC$. Here,
$\CC^*:=\CC-\{0\}$, and $\hat\CC:=\CC\cup\{\infty\}$ denotes the
extended complex plane. We briefly discuss these properties
globally in terms of (linear fractional) M\"obius transformations
of the extended complex plane $\hat\CC\ni z$, in terms of the
deformation mapping of the $\CC\PP^1-$related Bloch sphere, as
well as locally in terms of the Fubini-Study metric.

An arbitrary state vector $|\psi\ra\in\cH=\CC^2$ can be represented
as\footnote{A relation of the type $\psi \cong
\left(1,z\right)^T\CC$ denotes the equivalence of $\psi$ to a point
of the projective space $\CC\PP^1$ represented by its equivalence
class $\left(1,z\right)^T\CC$. See, e.g. \cite{moebius-1}. The full
$\CC$ is allowed, in general, because the point $\{0\}$ can be
combined in a controlled way with $z=\infty $ to pass onto the other
affine chart with $w=1/z $ as coordinate.}
\be{q3}
|\psi\rangle=\cos(\t)|0\rangle+e^{i\phi}\sin(\t)|1\rangle=\left(%
\begin{array}{c}
  \cos(\t) \\
   e^{i\phi}\sin(\t)\\
\end{array}%
\right)\cong \left(%
\begin{array}{c}
  1 \\
  z \\
\end{array}%
\right)\CC
\ee
with $z=e^{i\phi}\tan(\t)\in\hat \CC$ as coordinate of the
extended complex plane $\hat\CC$. A linear transformation
\be{q4}
S:\ |\psi\ra\mapsto |\psi'\ra=S|\psi\ra,\qquad S=\left(%
\begin{array}{cc}
  A & B \\
  C & D \\
\end{array}%
\right)\in SL(2,\CC)
\ee
acts then as linear fractional (M\"obius) transformation $M(2)\cong
PSL(2,\CC)$ \cite{moebius-1} (automorphism Aut$(\hat\CC)$) on
$z\in\hat\CC$
\be{q5}
S:\ z\mapsto z'=f(z):=\frac{Dz+C}{Bz+A}\,.
\ee
Apart from the their decomposition properties (translation,
rotation, dilation, inversion) M\"obius transformations are
classified by their type and fixed points $z=f(z)$. For $S\in
SL(2,\CC)$ the type is given by $T(S):=[\tr(S)]^2$ as $T=4$
--- parabolic, $T\in[0,4)$ --- elliptic, $T\in(4,\infty)$
--- hyperbolic and $\CC\ni T\not\in[0,4]$ --- loxodromic \cite{encyc-jap1}.
For the similarity transformation $\rho$ in \rf{q2} it holds
$T(\rho)=4\cosh^2(\b/2)$ so that for $\b\neq 0$ it is hyperbolic and
in the trivial case $\b=0$, $\rho=I_2$ --- parabolic. All
non-trivial transformations $\rho(\b)$ have the same pair of fixed
points $z_e\equiv z_\pm=\pm i$ independently of the value of
$\b\neq0$. Comparison with \rf{h3} shows that the fixed point states
$|I_\pm\ra:=(1,\pm i)^T\CC^*$ correspond to the eigenvectors at the
exceptional points of $H(\a=\mp \frac\pi 2+2N\pi)$, $N\in\ZZ$. A
point $z=z_\e+\Delta$, $|\Delta|\ll 1$ close to a fixed point maps
as
\be{q6}
z\mapsto z'=f(z_\e+\Delta)\approx
f(z_\e)+f'(z_\e)\Delta=:z_\e+\Delta'
\ee
so that from $f'(z_\e)=\exp(-\e\b)$ one finds a distance dilation
$f'>1$ for $\e\b<0 $ and a contraction $f'<1$ for $\e\b>0 $. Hence,
for $\b>0$ ($\b<0$) the fixed point $z_+$ ($z_-$) acts as attractor
and $z_-$ ($z_+$) as repellor (see, e.g. \cite{attract-repell}).

Closely related to the distances on $\hat\CC\ni z$ is the
Fubini-Study metric (see, e.g. \cite{nakahara}) on
$\cP(\cH)=\CC\PP^1$. In terms of the affine coordinate $z$ this
metric reads
\be{q6a}
ds^2=\frac{2dz d z^*}{\left(1+|z|^2\right)^2}=: g(z,z^*)dz d
z^*\,.
\ee
Under the mapping $\rho:\ \cP(\cH)\to\cP(\tilde \cH)$ it
transforms into $d\tilde s^2=g(z',z'^*)dz'dz'^*$ with $z'$ given
by \rf{q5} and $S=\rho$. In terms of the original affine
coordinate $z$ it takes the form
\be{q6b}
d\tilde s^2=\frac{2dz
dz^*}{\left[\cosh(\b)(1+|z|^2)+i\sinh(\b)(z^* -z)\right]^2}
\ee
and coincides with the $\eta-$deformed Fubini-Study metric of the
Hilbert space $(\cH_\eta,\la.|.\ra_\eta)$ (see \rf{a5} in the
Appendix). For the fixed point vicinities with $z\approx z_\e=\e
i$ the metric \rf{q6b} reproduces (already in zeroth-order
approximation) the typical contraction/dilation
(attractor/repellor) behavior $d\tilde s^2\approx 2e^{-\e\b}dz
dz^*$ found via \rf{q6}.

\begin{figure}[th]
\begin{center}
\begin{minipage}{0.51\textwidth}
\includegraphics[angle=0, width=0.9\textwidth]{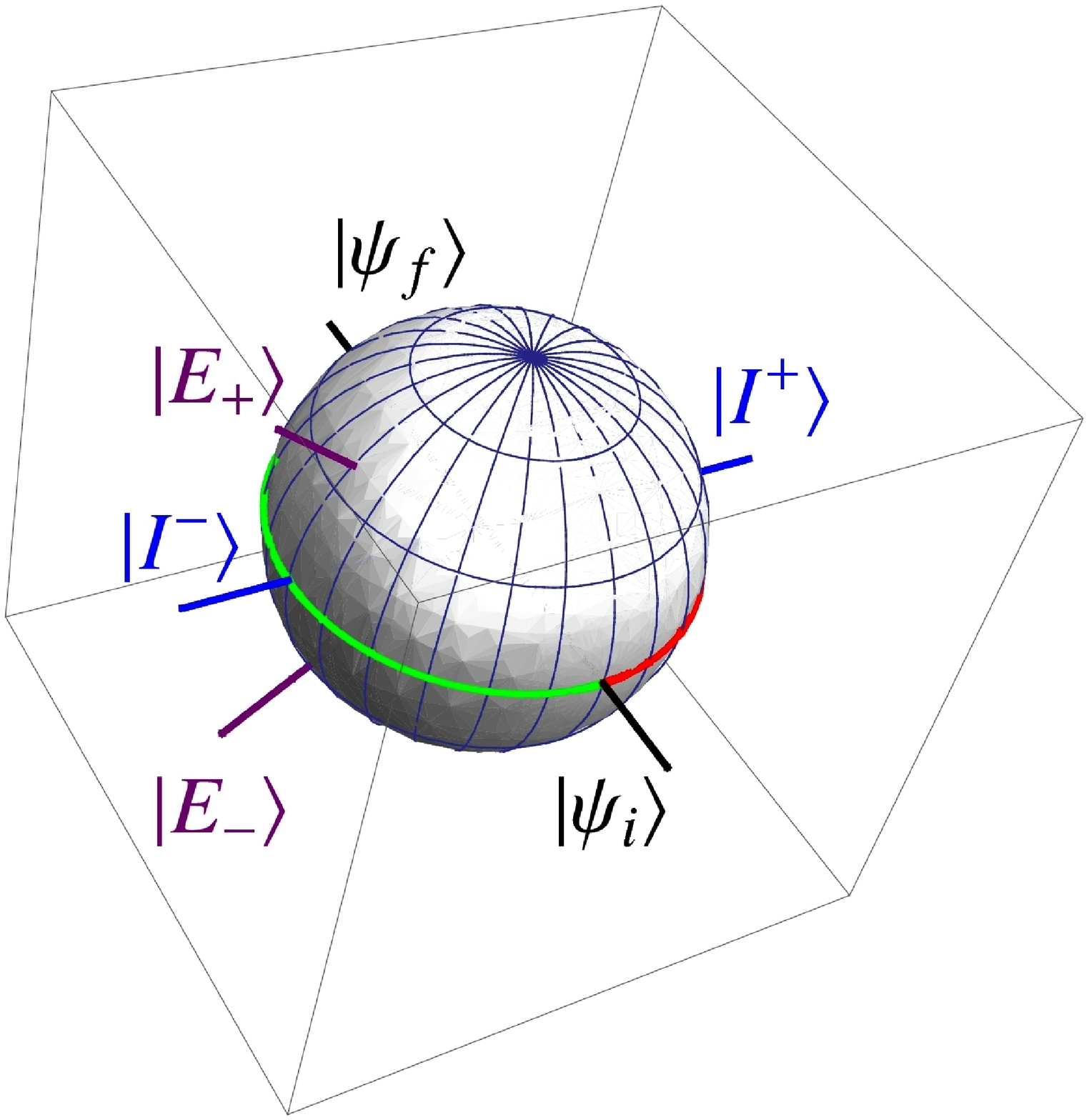}
\end{minipage}
\begin{minipage}{0.41\textwidth}
\includegraphics[angle=0, width=0.9\textwidth]{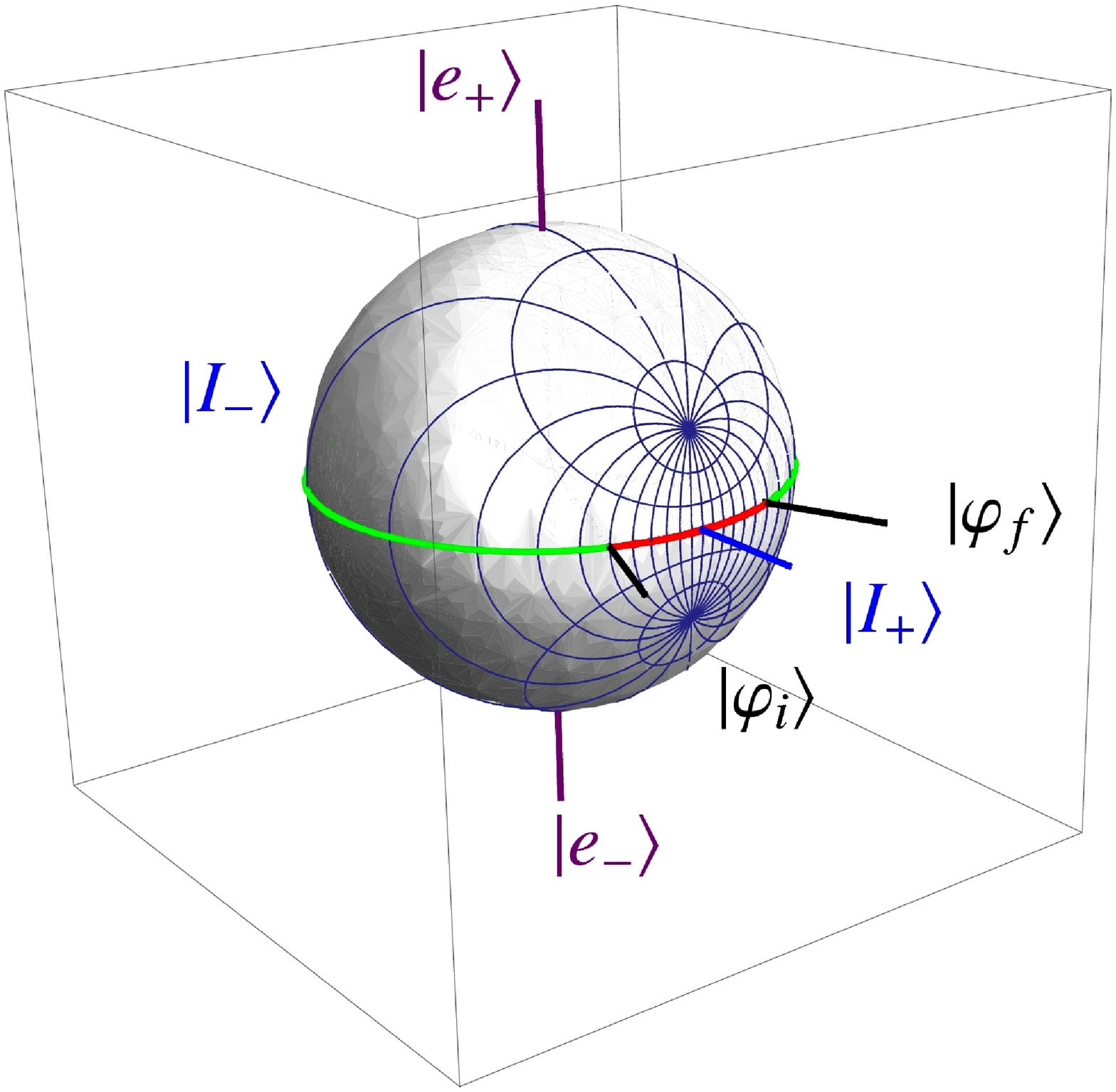}
\end{minipage}
\end{center}
\caption{\label{fig1}\small The transformation $\rho$ maps the
initial and final states $|\psi_{i,f}\ra\in\cH$ as well as the
energy eigenstates $|E_\pm\ra\in\cH$ into
$|\phi_{i,f}\ra,|e_\pm\ra\in\tilde\cH$, respectively, and leaves the
EP-related fixed point states $|I_\pm\ra$ invariant. The
contraction/dilation properties of the evolution paths (high-lighted
red/green curves) are defined by their location relative to the
originally non-orthogonal energy eigenstates $|E_\pm\ra$.}
\end{figure}

A next piece of information can be gained by considering the
mapping $\rho$ globally as automorphism of the Bloch sphere. The
Bloch sphere representation of a quantum state $\psi$ is given by
the correspondence $\psi\in\CC^2\to \CC\PP^1\cong S^2\subset
\RR^3$ which for a state parametrization \rf{q3} has the
form\footnote{We note that the natural distance between two states
$|\psi_1\ra, |\psi_2\ra\in\cH$ is given by the corresponding
angular distance on the Bloch sphere
dist${}_{\cH}(\psi_1,\psi_2):=2\arccos(|\la\psi_2|\psi_1\ra|)$ and
that two states are orthogonal when they are antipodal on this
sphere.}
\be{q7}
x=\sin(2\t)\cos(\phi),\qquad y=\sin(2\t)\sin(\phi),\qquad
z=\cos(2\t)\,.
\ee
We use this representation together with the projective mapping of
an arbitrary non-normalized state vector $\psi\in\CC^2$
\be{q8}
\psi=\left(%
\begin{array}{c}
  a \\
  b \\
\end{array}%
\right)\cong \left(%
\begin{array}{c}
  \cos(\t) \\
  e^{i\phi}\sin(\t) \\
\end{array}%
\right)\CC^*
\ee
and the easily derived relations
\be{q9}
\phi=\arg(b)-\arg(a),\qquad
\cos(2\t)=\frac{|a|^2-|b|^2}{|a|^2+|b|^2},\qquad
\sin(2\t)=\frac{2|a||b|}{|a|^2+|b|^2}
\ee
to analyze the $\rho-$induced transformations graphically. The
corresponding plots in Fig. \ref{fig1} demonstrate the global
deformations induced by $\rho$.
Clearly visible is the relative position of the states in the
Hilbert spaces. In the space $(\cH,\la.|.\ra)$ the eigenstates
(-vectors) $|E_\pm\ra$ of the non-Hermitian $\cP\cT-$symmetric
Hamiltonian $H$ are non-orthogonal (non-antipodal), whereas the
initial and final eigenstates $|\psi_i\ra,|\psi_f\ra$ of the
Hermitian spin operator $\sigma_z$ are orthogonal (antipodal). The
mapping $\rho: \cH\to \tilde\cH$ acts in such a way that it
transforms $|E_\pm\ra$ into the states $|e_\pm\ra$ which are
orthogonal in $(\tilde\cH,\la.|.\ra)$. Simultaneously, it transforms
$|\psi_i\ra,|\psi_f\ra$ into the non-orthogonal
$|\phi_i\ra=\rho|\psi_i\ra$ and $|\phi_f\ra=\rho|\psi_f\ra$ dilating
or contracting in this way the distance
dist${}_{\cH}(|\psi_i\ra,|\psi_f\ra)=\pi$ into
dist${}_{\tilde\cH}(|\phi_i\ra,|\phi_f\ra)\gtrless\pi$.  The
antipodal fixed point states $|I_\pm\ra$ remain invariant under
$\rho$. The states $|E_\pm\ra$ are located on a big circle passing
through the fixed points $|I_\pm\ra$, and $|\psi_i\ra,|\psi_f\ra$ on
another $\pi/2-$rotated big circle through $|I_\pm\ra$. Under the
transformation $\rho$ all but the fixed point states are moved along
these big circles away from the repellor fixed point and toward the
attractor fixed point.

In $\tilde\cH$ the evolution between the states
$|\phi_i\ra=\rho|\psi_i\ra$ and $|\phi_f\ra=\rho|\psi_f\ra$ is
governed by the unitary transformation $u(t)=e^{-ith}$ with
Hermitian Hamiltonian $h$. This unitary transformation corresponds
to the usual rigid rotation of the Bloch sphere \cite{brody-1}
(elliptic type M\"obius transformation) with the two mapped energy
eigenstates $|e_\pm\ra=\rho|E_\pm\ra$ as antipodal transformation
fixed points. In \cite{GRS-EP-JPA} it has been shown that the
vanishing-passage-time solution of the brachistochrone problem of
\cite{cmb-brach} corresponds to an EP-limit with coalescing energy
eigenstates $|E_+\ra\to|E_-\ra$. The mapping $\rho$
`orthogonalizes' them into $|e_\pm\ra$ but simultaneously
transforms the orthogonal $|\psi_{i,f}\ra$ into coalescing
$|\phi_i\ra\to|\phi_f\ra$ and induces a corresponding vanishing
distance dist${}_{\tilde\cH}(|\phi_i\ra,|\phi_f\ra)\to 0$. The
evolution type is not affected by this equivalence, i. e. the
transformation $U(t):\ \cH\to\cH$ remains pseudo-unitary with
regard to $(\cH_\eta,\la.|.\ra_\eta)$ and $u(t):\
\tilde\cH\to\tilde\cH$ unitary with regard to $\tilde\cH$. For
$u(t): |\phi_i\ra\mapsto |\phi_f\ra$ the Anandan-Aharonov lower
bound \cite{anandan-aharonov-prl-1990} on the passage time remains
valid.

Finally, we note that the hyperbolic type M\"obius transformation
on the Bloch sphere with its two transformation fixed points and
the distance contraction and dilation mechanism is a generic
projective transformation which in relativistic physical systems
induces the well known aberration effect \cite{rindler} of
shifting the positions of far stars toward the direction of motion
of the relativistically moving observer.

\section{Concluding remarks\label{conclu}}

In the present paper we have interpreted the $\cP\cT-$symmetric
brachistochrone setup of \cite{cmb-brach} as a quantum system
consisting of a non-Hermitian $\cP\cT-$symmetric component and a
Hermitian component simultaneously. This interpretation allowed us
to formulate a general recipe for the construction of partially
$\cP\cT-$symmetric quantum systems which are not 1:1 equivalent to
purely Hermitian systems. Using a strong structural analogy with
the reference frames for inertial observers in special relativity
we associated $\cP\cT-$symmetric models in different
representations with corresponding measurement frames. We showed
that operators which are Dirac Hermitian are connected with
non-Dirac-Hermitian operators in another frame. The probabilistic
content of the models is frame-independent. With the help of a
geometric analysis of the equivalence mapping between mutually
$\cP\cT-$symmetric and Hermitian operators the compatibility of
the vanishing passage-time solution with the Anandan-Aharonov
lower bound \cite{anandan-aharonov-prl-1990} for Hermitian system
has been demonstrated.

\section*{Acknowledgements} We thank Lajos Di\'osi for drawing our
attention to his recent monograph \cite{diosi}. The work has been
supported by the German Research Foundation DFG, grant GE 682/12-3,
(UG), by the Saxonian Ministry of Science (grant
4-7531.50-04-844-07/5), (BFS) as well as by the grants
RFBR-06-02-16719, SS-871.2008.2 (BFS). BFS also thanks the Research
Center Dresden-Rossendorf for hospitality during his stay in
Dresden.
\appendix

\section{$\eta-$deformed Fubini-Study metric}
The Fubini-Study metric \cite{nakahara} on a standard QM-related
projective Hilbert space $\PP(\tilde \cH)=\CC\PP^N$ is given in
terms of state vectors $|\phi\ra\in \tilde\cH=\CC^{N+1}$ as
\be{a1}
d\tilde s^2=2\frac{\la\phi|\phi\ra\la d\phi|d\phi\ra-\la
d\phi|\phi\ra\la\phi|d\phi\ra}{\la\phi|\phi\ra^2}\,.
\ee
When the states $|\phi\ra$ are the result of a linear invertible
mapping $\rho:\ |\psi\ra\mapsto |\phi\ra=\rho|\psi\ra$ with
$\rho^\dd\rho=\eta$ then for $|\psi\ra\in \cH=\CC^{N+1}$ the metric
\rf{a1} becomes `$\eta-$deformed' (see also \cite{ali-brach-2007})
\be{a2}
d\tilde s^2=2\frac{\la\psi|\eta|\psi\ra\la d\psi|\eta|d\psi\ra-\la
d\psi|\eta|\psi\ra\la\psi|\eta|d\psi\ra}{\la\psi|\eta|\psi\ra^2}\,.
\ee
For the affine chart $U_0\ni |\psi\ra=(1,z_1,\ldots,z_N)^T=:
(1,z)^T$ of the projective space $\CC\PP^N\supset U_0$ one sets for
convenience
\be{a3}
\eta=\left(%
\begin{array}{cc}
  a & c^\dd \\
  c & D \\
\end{array}%
\right),\qquad a\in\RR,\quad c\in\CC^N,\quad D\in\CC^{N\times N}
\ee
and finds (due to $\eta=\eta^\dd, \ D=D^\dd$)
\be{a4}
d\tilde s^2=2\frac{dz^\dd[qD-(c+Dz)\otimes
(c+Dz)^\dd]dz}{q^2},\quad q:=a+c^\dd z+z^\dd c+z^\dd D z\in\RR\,.
\ee
In the case of $\det (\eta)=1$ and $|\psi\ra\in\CC\PP^1$ this
reduces via $D\in\RR$ to
\be{a5}
d\tilde s^2=\frac{2dz^*dz}{\left[a+c^*z+cz^*+D|z|^2\right]^2}\,.
\ee


\begin{thebibliography}{99}

\bibitem{BB-PTSQM-1}C. M. Bender and S. Boettcher, "Real Spectra in Non-Hermitian Hamiltonians Having $\cP\cT$
Symmetry", Phys. Rev. Lett. {\bf 80}, (1998), 5243,
physics/9712001.

\bibitem{cmb-rev} C. M. Bender, "Making sense of non-Hermitian
Hamiltonians", Rep. Prog. Phys. {\bf 70}, (2007), 947-1018,
hep-th/0703096.

\bibitem{ali-herm-1} A. Mostafazadeh, "Pseudo-Hermiticity versus PT-symmetry II:
A complete characterization of non-Hermitian Hamiltonians with a
real spectrum", J. Math. Phys. {\bf 43}, (2002) 2814-2816,
math-ph/0110016.

\bibitem{cmb-nonlocal}C. M. Bender, D. C. Brody, H. F. Jones, "Extension of PT-symmetric quantum mechanics to quantum field theory
with cubic interaction", Phys. Rev. D {\bf 70}, (2004), 025001;
Erratum-ibid. D {\bf 71}, (2005), 049901, hep-th/0402183.

\bibitem{ali-nonlocal} A. Mostafazadeh, "Exact PT-symmetry is equivalent to
Hermiticity", J. Phys. A {\bf 36}, (2003), 7081-7092,
quant-ph/0304080; "PT-symmetric cubic anharmonic oscillator as a
physical model", J. Phys. A {\bf 38}, (2005), 6557-6570;
Erratum-ibid. A {\bf 38}, (2005), 8185, quant-ph/0411137.

\bibitem{cmb-brach} C. M. Bender, D. C. Brody, H. F. Jones and B.
K. Meister, "Faster than Hermitian quantum mechanics", Phys. Rev.
Lett. {\bf 98}, (2007), 040403, quant-ph/0609032.


\bibitem{ali-brach-2007}A. Mostafazadeh,
"Quantum brachistochrone problem and the geometry of the state
space in pseudo-Hermitian quantum mechanics", Phys. Rev. Lett.
{\bf 99}, 130502, (2007), quant-ph/0706.3844.

\bibitem{GRS-EP-JPA} U. G\"unther, I. Rotter and B. Samsonov,
"Projective Hilbert space structures at exceptional points", J.
Phys. A {\bf 40}, (2007), 8815-8833, math-ph/0704.1291.



\bibitem{Holevo}
A.S. Holevo, {\it Statistical structure of quantum theory}
(Berlin: Springer-Verlag ,2001); A.S. Holevo {\it Probabilistic
and statistical aspects of quantum theory}, (Moscow: Nauka 1980).



\bibitem{Peres} A. Peres, {\it Quantum theory: concepts and
methods}, (Kluwer, Dordrecht, 1993).

\bibitem{Nielsen} M.A. Nielsen and I.L. Chuang, {\it Quantum computation and quantum information},
(Cambridge University Press, Cambridge, 2000).

\bibitem{QI}
C.W. Helstrom, {\it Quantum detection and
estimation theory}, (Academic: New York, 1976), pp74-83.\\
D.B. Osteyee and I.J. Good, {\it Information, weight of evidence,
the singularity between probability measures, and signal detection},
Lect. Notes in Math. {\bf 376} (New York: Springer-Verlag,
1974).\\
A. Peres, Found. Phys. {\bf 20} 1441
(1990).\\
J. Bergou, U. Herzog and M. Hillery, {\it Discrimination of quantum
states}, Lect. Notes Phys. {\bf 649}, 417 (Springer: Berlin,
2004).\\
J.A. Bergou J. Phys.: Conf. Ser. {\bf 84}, 012001 (2007).

\bibitem{diosi} L. Di\'osi, {\it A short course in quantum information
theory}, Lect. Notes Phys. 713, (Springer, Berlin, 2007).


\bibitem{krein-pts} H. Langer and C. Tretter,
"A Krein space approach to PT-symmetry", Czech. J. Phys. {\bf 54}, 1113
(2004);
S. Albeverio and S. Kuzhel, "Pseudo-Hermiticity and theory
of singular perturbations", Lett. Math. Phys. {\bf 67}, (2004),
223-238;
U. G\"unther, F. Stefani and M. Znojil,
"MHD $\a^2-$dynamo,
Squire equation and PT-symmetric interpolation between square well
and harmonic oscillator", J. Math. Phys. {\bf 46}, (2005), 063504,
math-ph/0501069.

\bibitem{azizov}T. Ya. Azizov and I.S. Iokhvidov,
{\it Linear operators in spaces with an indefinite metric}
(Wiley-Interscience, New York, 1989).

\bibitem{L2} A. Dijksma and H. Langer, {\it Operator theory and ordinary differential operators}, in
A. B\"ottcher (ed.) {\it et al.}, {\it Lectures on operator theory
and its applications}, Providence, RI: Am. Math. Soc., Fields
Institute Monographs, {\bf Vol. 3},  75 (1996).

\bibitem{hahne} F. G. Scholtz, H. B. Geyer and F. J. W. Hahne,
"Quasi-Hermitian operators in quantum mechanics and the
variational principle", Ann. Phys. {\bf 213}, (1992), 74-101.

\bibitem{ali-pseudo-herm} A. Mostafazadeh,
"Pseudo-Hermiticity versus PT symmetry: The necessary condition
for the reality of the spectrum of a non-Hermitian Hamiltonian",
J. Math. Phys. {\bf 43}, (2002), 205-214, math-ph/0107001.

\bibitem{ali-eta-JMP}A. Mostafazadeh, "Pseudo-Hermiticity and generalized PT- and CPT-symmetries",
J. Math. Phys. {\bf 44}, (2003), 974-989, math-ph/0209018.

\bibitem{ryder} L. H. Ryder, {\it Quantum field theory}, (Cambridge: Cambridge Univ. Press, 1985).

\bibitem{davis} E. B. Davis, IEE Trans. Inform. Theory {\bf
IT-24}, 596 (1978).

\bibitem{Our2}B.F. Samsonov and U. G\"unther,
(in preparation).

\bibitem{anandan-aharonov-prl-1990} J. Ananandan and Y. Aharonov,
"Geometry of quantum evolution", Phys. Rev. Lett. {\bf 65},
(1990), 1697-1700.

\bibitem{giri-brach}P. R. Giri, "Lower bound of minimal time evolution in quantum
mechanics", Int. J. Theor. Phys. {\bf 47}, (2008), 2095-2100,
quant-ph/0706.3653.

\bibitem{brody-1} D. J. Brody, "Elementary derivation for passage
time", J. Phys. A {\bf 36}, (2003), 5587-5593, quant-ph/0302067; D.
J. Brody and D. W. Hook, "On optimum Hamiltonians for state
transformations", J. Phys. A {\bf 39}, (2006), L167-L170,
quant-ph/0601109.



\bibitem{moebius-1} U. Hertrich-Jeromin, {\it Introduction to M\"obius differential geometry.}, (Cambridge: Cambridge Univ. Press, 2003).
\bibitem{encyc-jap1} K. Ito, {\it Encyclopedic dictionary of mathematics},
(MIT press, Cambridge (MA), 1993).
\bibitem{attract-repell} J. Guckenheimer and P. Holmes, {\it Nonlinear oscillations, dynamical systems,
and bifurcations of vector fields}, (New York: Springer, 1983).
\bibitem{nakahara} M. Nakahara, {\it Geometry, topology and
physics}, (IOP Publishing, Bristol, 1990).


\bibitem{rindler} W. Rindler, {\it Relativity: special, general, and
cosmological}, (Oxford, Oxford Univ. Press, 2007).











\end{thebibliography}
\end{document}